# Tunable laser and photocurrents from linear atomic C chains

Zheng-Zhe Lin*

*School of Physics and Optoelectronic Engineering, Xidian University, Xi'an 710071, China*

*Corresponding Author. E-mail address: linzhengzhe@hotmail.com

**Abstract** - By a tight-binding model, the interaction between linear atomic C chains (LACCs) and short laser pulses was investigated. LACCs were proposed to be used as a medium of laser whose wavelength can be continuously tuned in a range of 321~785 nm. The lifetime of conduction band bottom is about 1.9~2.3 ns. The electrons pumped into conduction band will quickly fall to the conduction band bottom by electron-phonon interactions in a time of ps. The above results indicate that LACCs are suitable for laser medium. By $\omega+2\omega$ dichromatic laser pulses, photocurrents can be generated in LACCs, which can be applied as light-controlled signals.

## I. Introduction

People have been pursuing low-dimensional materials over the past decade for the design of nanodevices. In 1930's, Landau and Peierls argued that low-dimensional crystals were thermodynamically unstable and could not exist [1, 2]. However, one-dimensional monatomic gold chains have been successfully prepared by pulling two contacted atom-sized junctions [3, 4], and similar technique was used for preparing copper, aluminum and platinum chains [5]. In recent years, the successful preparation of single-layer graphene [6-9], silicene [10-12] and $MoS_2$ [13, 14] *et al.* has strongly challenged Landau's theory again. Following the preparation of graphene, free-standing linear atomic C chains (LACCs) were carved out from single-layer graphene by a high-energy electron beam [15] or unraveled from sharp carbon specimens [16]. Moreover, other techniques for preparing LACCs were also theoretically proposed [17, 18]. Due to the successful preparation of one-dimensional structures, people also paid attention to their physical properties. For example, LACCs capped with magnetic transition-metal atoms were found ferromagnetic or antiferromagnetic, and they can be used as spin-valve when connected to appropriate metallic electrodes [19-21]. Moreover, the structural, electronic, and magnetic properties of 3d transition metal monatomic chains have been systematically studied [22]. In our recent work, a statistical mechanical model beyond Landau's theory was provided to predict the stability of nanosystems [23]. By our model, LACCs were predicted to be thermally and chemically stable [23, 24]. We also investigated methods to modulate the electronic properties of LACCs [25], and proposed doped LACCs for detecting single gas molecule [26]. However, the optical properties of LACCs haven't been widely studied.

In this work, the interaction between LACCs and short laser pulses was investigated, and LACCs were proposed to be used as medium of continuously tunable laser. By mechanical pulling, its wavelength can be continuously tuned in a range of 321~785 nm. In the elastic deformation range of LACCs, the lifetime of conduction band bottom is about 1.9~2.3 ns, which is long enough as a laser medium. The electrons pumped to the conduction band will quickly fall to the conduction band bottom by electron-phonon interactions in a time of ps. By ω+2ω dichromatic laser pulses, photocurrents can be generated in LACCs when $\hbar\omega$ is far off resonance or near-resonance. Such photocurrents can be applied as light-controlled signals.

**II. Tight-binding model**

In this section, we briefly introduce the tight-binding model of LACC. In a LACC along the $x$-axis, the $sp_x$, $p_y$, $p_z$ orbitals on C atoms (Fig. 1(a)) constitute one $\sigma$ and two degenerate $\pi$ bands. The $\pi$ bands possess an energy range around the Fermi level. For a LACC of $N$ atoms, we used $|1>, |2>, \ldots |N>$ to denote $p_y$ (or $p_z$) orbitals on every atom. With the overlaps between neighboring orbitals ignored, they are treated as orthogonal. For the C atoms locating at $\mathbf{x_1}, \mathbf{x_2} \ldots \mathbf{x_N}$, the electronic Hamiltonian

reads

$$H_0 = \sum_j \varepsilon_0 c_j^+ c_j - \sum_j t(|\mathbf{x_{j+1}} - \mathbf{x_j}|)(c_j^+ c_{j+1} + c_{j+1}^+ c_j), \qquad (1)$$

where $\varepsilon_0=0$ is the energy of $p_y$ (or $p_z$) orbital on an isolated C atom and $t(x)$ is the hopping parameter. The operator $c_j$ ($c_j^+$) annihilates (creates) a fermion on the $i_{th}$ atom and satisfies the anticommutation relations $\{c_{n1} c_{n2}^+\} = \delta_{n1,n2}$. For the hopping parameter, an empirical form $t(x)=A\exp(-\alpha x)$ was used.

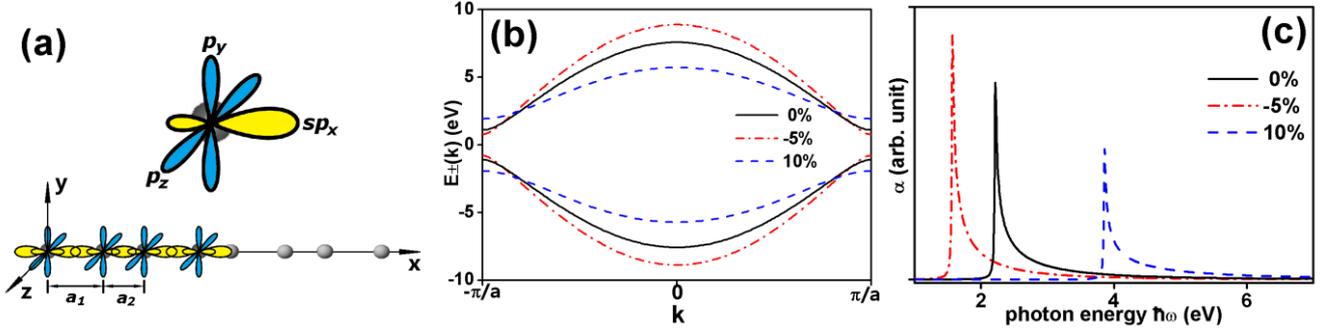

Fig. 1 (a) The structure of LACC and $sp_x$, $p_y$ and $p_z$ orbitals on C atoms. (b) The $E\sim k$ relation of LACC calculated by Eq. (5) with strain $s=0$ (solid lines), -5% (dot-dashed lines) and 10% (dashed lines) applied. (c) The photon absorption coefficient $\alpha$ with $\mathbf{A_0}$ parallel to LACC and strain $s=0$ (solid lines), -5% (dot-dashed lines) and 10% (dashed lines) applied.

Due to the Peierls distortion [27], the structure of LACC presents short and long C-C bond length alternation, i.e. $|\mathbf{x_{2n}}-\mathbf{x_{2n-1}}|=a_1$ and $|\mathbf{x_{2n+1}}-\mathbf{x_{2n}}|=a_2$. The lattice constant reads $a=a_1+a_2$. For a given wave vector $k$, Bloch states of the atomic orbitals read

$$|k1> = \frac{1}{\sqrt{N}} \sum_n |2n-1> \exp(ikna) \qquad (2)$$

$$|k2> = \frac{1}{\sqrt{N}} \sum_n |2n> \exp(ik(na+a_1)), \qquad (3)$$

and the $\pi$ band states reads

$$|k_\pm> = \frac{1}{\sqrt{2}} (|k1> - \frac{E_\pm(k)}{t(a_1)\exp(ika_1)+t(a_2)\exp(-ika_2)} |k2>), \qquad (4)$$

where

$$E_\pm(k) = \pm\sqrt{t^2(a_1)+t^2(a_2)+2t(a_1)t(a_2)\cos ka} \qquad (5)$$

is the band energy. By Eq. (5), the $\pi$ band splits into valence (VB) and conduction band (CB) with a direct band gap $E_g=E_+(\pm\pi/a)-E_-(\pm\pi/a)=2[t(a_1)-t(a_2)]$.

In the following text, the coefficients $A$ and $\alpha$ for $t(x)=A\exp(-\alpha x)$ was adjusted to meet previous *ab initio* calculation. Recently, Ming *et al.* performed hybrid density functional calculations for LACC with adjusted mixing parameter [28]. The C-C bond lengths and band gap were reported to be $a_1$=1.219 Å, $a_2$=1.332 Å and $E_g$=2.21 eV. In the range of elastic deformation (corresponding to a strain of $s$=-5%~10%), $E_g$ changes in a range of 1.58~3.86 eV. By the reported bond lengths, we derived $a_1=(1.332+1.837s+2.233s^2)$ Å and $a_2=(1.219+0.714s-2.233s^2)$ Å, and the lattice constant reads $a=a_1+a_2=(2.551+2.551s)$ Å. Then, the coefficients of $t(x)$ was fitted as $A$=103.28 eV and $\alpha$=2.6 Å$^{-1}$. By the fitted parameters, the $E$~$k$ profiles for strain $s$=0, -5% and 10% were plotted in Fig. 1(b), reproducing $\pi$ band structures close to the data in Ref. [28].

**III. Absorption spectrum**

The interaction between the LACC and photons can be described by an interaction Hamiltonian

$$H'=\frac{e\mathbf{A}\cdot\mathbf{p}}{m}=\sum_{j,l}\frac{e\mathbf{A}\cdot\mathbf{p}_{jl}c_j^+c_l}{m}, \qquad (6)$$

where $\mathbf{A}$ and $\mathbf{p}$ are the vector potential and electronic momentum, respectively. Here,

$$\begin{aligned}\mathbf{p}_{jl} &=<j|\mathbf{p}|l>=-i\hbar\int\varphi^*(\mathbf{x}-\mathbf{x}_j)\nabla\varphi(\mathbf{x}-\mathbf{x}_l)d^3\mathbf{x}\\ &=i\hbar\nabla S(\mathbf{r}=\mathbf{x}_l-\mathbf{x}_j)\end{aligned}, \qquad (7)$$

where $\varphi(\mathbf{x}\text{-}\mathbf{x}_l)$ is the $p$ orbital locating on the $l_{th}$ atom and

$$S(\mathbf{r})=\int\varphi^*(\mathbf{x})\varphi(\mathbf{x}-\mathbf{r})d^3\mathbf{x} \qquad (8)$$

is the overlap integral. We roughly set $S(\mathbf{r})=\exp(-\beta r^2)$ and Eq. (7) reads

$$\mathbf{p}_{jl}=2i\hbar(\mathbf{x}_l-\mathbf{x}_j)\beta\exp(-\beta|\mathbf{x}_l-\mathbf{x}_j|^2). \qquad (9)$$

To investigate photon absorptivity of LACC, we considered monochromatic photons $\mathbf{A}=\mathbf{A}_0\cos(\omega t)$. By Fermi's golden rule, the transition rate from the state $|k_-\rangle$ in VB to $|k_+\rangle$ in CB reads

$$W(k) = \frac{\pi e}{m\hbar} \delta(E_+ - E_- - \hbar\omega) |< k_- | \mathbf{A_0} \cdot \mathbf{p} | k_+ >|^2$$
$$= \frac{\pi e}{m\hbar} \delta(E_+ - E_- - \hbar\omega) \sum_{j,l} |\mathbf{A_0} \cdot \mathbf{p_{jl}}|^2 |< k_- | c_j^+ c_l | k_+ >|^2 \quad , \tag{10}$$

and the absorption coefficient reads

$$\alpha \propto \int_{-\pi/a}^{\pi/a} W(k) dk$$
$$\propto \int_{-\pi/a}^{\pi/a} \delta(E_+ - E_- - \hbar\omega) (\sum_{j,l} |\mathbf{A_0} \cdot \mathbf{p_{jl}}|^2 |< k_- | c_j^+ c_l | k_+ >|^2) dk \quad . \tag{11}$$

When $\mathbf{A_0}$ is perpendicular to the LACC, it is easy to know that $\alpha=0$ since $\mathbf{A_0} \mathbf{p_{jl}}=0$. So, in the following text $\mathbf{A_0}$ parallel to LACC was concerned. By Eq. (9), the on-site matrix elements $\mathbf{A_0} \mathbf{p_{nn}}=0$, and for the nearest neighbor atoms we have

$$\mathbf{A_0} \cdot \mathbf{p_{2n-1\ 2n}} = 2i\hbar A_0 a_1 \exp(-\beta a_1^2)$$
$$\mathbf{A_0} \cdot \mathbf{p_{2n\ 2n+1}} = 2i\hbar A_0 a_2 \exp(-\beta a_2^2) \quad . \tag{12}$$

Other matrix elements are much smaller and then ignored. Then, the empirical parameter $\beta$ was adjusted to 1.0 Å$^{-2}$ to meet the *ab initio* calculation of absorption spectrum [28]. By this parameter, the calculated absorption spectrum of LACC with strain $s=0$, -5% and 10% were plotted in Fig. 1(c), showing a maximum at the transition from VB top to CB bottom. The strength of the maximum absorption decreases with increasing strain $s$.

**IV. Monochromatic laser pulse**

To investigate basic optical properties, the interaction between LACC and monochromatic laser pulses was simulated. The electric field reads

$$\mathbf{E} = \begin{cases} \mathbf{E_0} \exp(-9(t-t_0)^2/t_0^2) \sin(\omega t) & 0 \leq t < t_0 \\ \mathbf{E_0} \sin(\omega t) & t_0 \leq t \leq T - t_0 \\ \mathbf{E_0} \exp(-9(t-t_0+T)^2/t_0^2) \sin(\omega t) & T - t_0 < t \leq T \end{cases}, \tag{13}$$

where $\mathbf{E_0}$, $t$ and $T$ are the amplitude, time and duration, respectively. Gaussian profile with $t_0=300$ fs and $T=1000$ fs was used for the starting and finishing stages of laser pulse. Electric field $\mathbf{E}$ is parallel to LACC, with a maximum intensity of $10^9$ W/cm$^2$ (i.e. $E_0=8.68\times10^7$ V/m). The corresponding vector potential reads $\mathbf{A} = -\int_0^t \mathbf{E} dt$. As an example, the plot of **E**~$t$ for $\hbar\omega$=0.2 eV is shown in Fig. 2(a). For a given laser wave packet with a vector potential **A**=**A**($t$), by the interaction described by Eq. (6) the Schrödinger equation for the evolution of every electronic states was simulated by the Runge-Kutta 4$_{th}$ order method. Under the excitation of laser field, the electrons in VB are pumped to CB, and the population of the state $|k_+\rangle$ in CB reads

$$P(k) = 4\sum_{j}|\langle k_+ | \Psi_j \rangle|^2, \qquad (14)$$

where $|\Psi_j\rangle$ is the wavefunction of the $j_{th}$ electron. The factor 4 comes from two spin degrees of freedom and two degenerate $p_y$, $p_z$ bands.

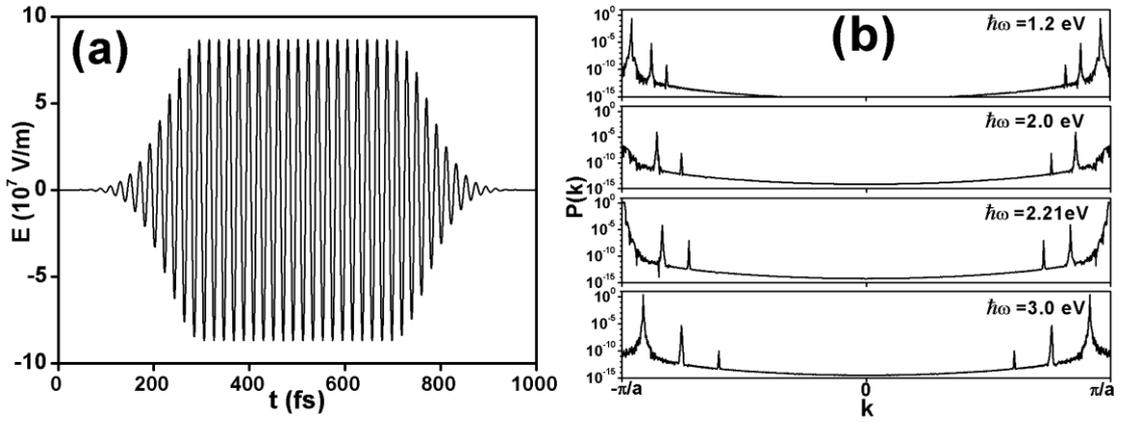

Fig. 2 (a) The plot of **E**~$t$ by Eq. (13) for $E_0$=8.68×10$^7$ V/m and $\hbar\omega$=0.2 eV. (b) The population $P(k)$ of the state $|k_+\rangle$ after the simulation by $E_0$=8.68×10$^7$ V/m and $\hbar\omega$=1.2, 2.0, 2.21 and 3.0 eV.

For LACC with atom number $N$=2000 and strain $s$=0, simulations of $\hbar\omega$=1.2, 2.0, 2.21 and 3.0 eV were performed, and $P(k)$ after are plotted in Fig. 2(b). When $\hbar\omega$ is smaller than the band gap $E_g$=2.21 eV, some peak values can be found in the $P(k)$~$k$ plot. For $\hbar\omega$=1.2 eV, the peaks locate at $k$=±0.96, ±0.88 and ±0.82 π/$a$ with $P(k)$=0.23, 7.5×10$^{-6}$ and 7.3×10$^{-10}$, respectively. At these $k$-points, corresponding energy difference $E_+(k)$-$E_-(k)$ between the state $|k_-\rangle$ in VB and $|k_+\rangle$ in CB are 2.4, 3.6 and 4.8 eV, respectively, i.e. 2$\hbar\omega$, 3$\hbar\omega$ and 4$\hbar\omega$. This result indicates the multi-photon absorption of LACC. For $\hbar\omega$=2.0 eV, the peaks locate at $k$=±0.86 and ±0.76 π/$a$ with $P(k)$=1.1×10$^{-4}$ and 7.0×10$^{-9}$, respectively, corresponding to the multi-photon absorption of 2$\hbar\omega$ and 3$\hbar\omega$. For $\hbar\omega$=2.21 eV which equals to the band gap $E_g$ of LACC, a strong peak $P(k)=1.73$ can be found at $k=\pm\pi/a$. And the other weaker peaks at $k=\pm 0.84$ and $\pm 0.73$ $\pi/a$ (with $P(k)=8.3\times 10^{-5}$ and $1.1\times 10^{-7}$) correspond to the multi-photon absorption of $2\hbar\omega$ and $3\hbar\omega$. For $\hbar\omega=3.0$ eV, the $P(k)$ peaks of the absorption of $\hbar\omega$, $2\hbar\omega$ and $3\hbar\omega$ (with $P(k)=2.6$, $4.9\times 10^{-6}$ and $1.0\times 10^{-10}$) locate at $k=\pm 0.91$, $\pm 0.76$ and $\pm 0.60$ $\pi/a$. The above result indicates that for $\hbar\omega>E_g$ single-photon and multi-photon absorption both exist in LACC.

**V. Tunable laser by LACC**

Since LACC has a large elastic deformation range ($s$=-5~10%) and its band gap $E_g$ varies with strain $s$, it was considered to be a medium for tunable laser [28, 29]. Such medium will be suitable if the state $|k_+\rangle$ in CB has an appropriate lifetime before falling back to $|k_-\rangle$ in VB. In CB, the energy of electrons at the $|k_+\rangle$ states will convert into atomic vibrational energy by electron-phonon interactions and finally the electrons fall to CB bottom before spontaneous radiation. By the deformation-potential approximation, the interaction Hamiltonian of electrons and longitudinal acoustical phonons reads

$$H'_{e-ph} = iDq\sqrt{\frac{\hbar}{2\mu L\Omega}}(C^+_{k+q}C_k b_q - C^+_{k-q}C_k b^+_q). \tag{15}$$

Here, the deformation-potential constant $D=dE_+(k=\pi/a)/ds$ denotes the variation of CB bottom energy along with strain $s$. $\mu$ and $L$ are the linear mass density and the length of LACC, respectively. $q$ and $\Omega$ are the wave vector and the angular frequency of phonon, respectively. The operator $b_q$ ($b^+_q$) annihilates (creates) a phonon with a wave vector $q$, and $C_k$ ($C^+_k$) annihilates (creates) an electron at the Bloch state $|k_+\rangle$. When existing $n(q)$ phonons with wave vector $q$, by Fermi's golden rule the rate of an electron at $|k_+\rangle$ releasing a phonon and transiting to $|(k-q)_+\rangle$ reads

$$U(q) = \frac{\pi D^2 q^2 (n(q)+1)}{\mu L\Omega}\delta(E_+(k-q)+\hbar\Omega - E_+(k)) \tag{16}$$

and the total decay rate of $|k_+\rangle$ reads

$$R = \frac{L}{2\pi} \int U(q)dq = \frac{D^2}{2\hbar\mu} \sum_j \frac{q_j^2(n(q_j)+1)}{\Omega(q_j)|d\Omega/dq|_{q_j}}, \tag{17}$$

where $q_j$ satisfies the energy conservation $E_+(k-q_j) + \hbar\Omega(q_j) - E_+(k) = 0$. For linear atomic chain, the dispersion relation of longitudinal acoustical phonon can be approximated as $\Omega \approx c_L|q|$, where $c_L = \sqrt{Y/\mu}$ is the longitudinal wave velocity near the $\Gamma$ point and $Y = \partial F/\partial s$ is the one-dimensional elastic modulus ($F$ is the tension in LACC). The maximum energy of released phonons is about $\hbar\Omega_{max} \approx \hbar c_L|\pi/a|$=0.24 eV. At temperature $T$, the average phonon number $n(q)=1/(\exp(\hbar\Omega/k_BT)-1) \approx k_BT/\hbar\Omega = k_BT/\hbar c_L|q|$. Then by Eq. (17) we have

$$\begin{aligned} R &\approx \frac{D^2}{2\hbar\mu} \sum_j \frac{|q_j|(\frac{k_BT}{\hbar c_L|q_j|}+1)}{c_L^2} \\ &\approx \frac{D^2}{2\hbar Y} \sum_j (|q_j| + \frac{k_BT}{\hbar\sqrt{Y/\mu}}) \\ &> \frac{D^2 k_B T}{2\hbar^2} \sqrt{\frac{\mu}{Y^3}} \end{aligned} \tag{18}$$

By *ab initio* calculation data in Ref. [28], we have $Y \approx 90$ eV/Å and $D \approx 7$ eV. By Eq. (18), it can be estimated that $R > 0.06$ fs$^{-1}$ for strain $s=0$. This result indicates that in 1 ps an electron in CB releases its energy in the form of phonons at least 60 times. Therefore, the electrons excited into CB will quickly lose energy and fall to CB bottom, i.e. $|k_+, k=\pm\pi/a\rangle$. Then, from the state $|k_+, k=\pm\pi/a\rangle$ the electrons return to VB top $|k_-, k=\pm\pi/a\rangle$ by spontaneous transition. The lifetime of $|k_+\rangle$ reads

$$\begin{aligned} \tau &= \frac{3\pi\hbar^2\varepsilon_0 mc^3}{e^2(E_+(k)-E_-(k))|\langle k_+|\mathbf{p}|k_-\rangle|^2} \\ &= \frac{3\pi\hbar^2\varepsilon_0 mc^3}{e^2(E_+(k)-E_-(k))|\sum_{j,l}\mathbf{p_{jl}}\langle k_+|c_j^+ c_l|k_-\rangle|^2} \end{aligned}. \tag{19}$$

For strain $s$=0, -5% and 10%, the result is plotted in Fig. 3(a). In CB, the minimum of $\tau$ locates at the band gap $k=\pm\pi/a$. In the vicinity of $k=0$, $\tau$ is much larger than at $k=\pm\pi/a$ and notably decreases with the elongation of LACC. At the band gap $k=\pm\pi/a$ the change of $\tau$ with the elongation is much smaller (the inset of Fig. 3(a)), which is 2.3~1.9 ns for $s$=-5%~10%. Such lifetime is enough for acting as a laser medium in the whole elastic deformation range, and the corresponding laser wavelength $\lambda=2\pi\hbar c/E_g$=785~321 nm. In this range, the laser wavelength can be tuned continuously by pulling the LACC.

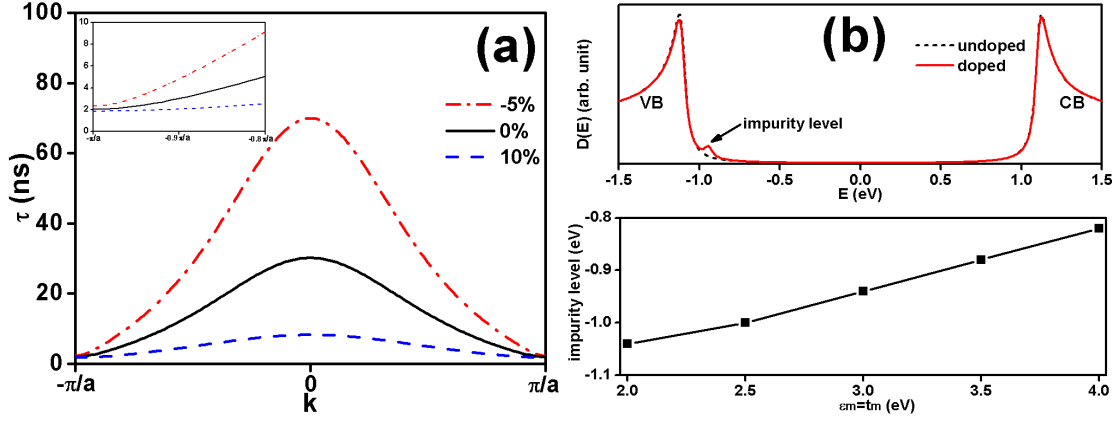

Fig. 3 (a) The lifetime $\tau$ of the electronic states in CB, with strain $s$=0 (solid lines), -5% (dot-dashed lines) and 10% (dashed lines) applied. (b) Upper: the density of states $D(E)$ for atom number $N$=1000, strain $s$=0 and $\varepsilon_m=t_m$=3.0 eV. Lower: for atom number $N$=1000, strain $s$=0 and $\varepsilon_m=t_m$, the energy of impurity level varies with $\varepsilon_m$ ($t_m$).

In semiconductor laser, doping technique has been widely used. By inserting acceptor impurity (e.g. B or Al atom) in LACC, the number of holes in VB is increased and the electron-hole recombination probability can be enhanced. To investigate the effect of doping on the energy band of LACC, we considered that the $m_{th}$ atom is substituted by a dopant, and the model Hamiltonian reads

$$H_{0m} = \sum_{j\neq m}\varepsilon_0 c_j^+ c_j - \sum_{j\neq m-1,m} t(|\mathbf{x_{j+1}}-\mathbf{x_j}|)(c_j^+ c_{j+1}+c_{j+1}^+ c_j) \\ + (\varepsilon_0+\varepsilon_m)c_m^+ c_m - \sum_{j=m-1,m}(t+t_m)(|\mathbf{x_{j+1}}-\mathbf{x_j}|)(c_j^+ c_{j+1}+c_{j+1}^+ c_j)$$ (20)

For acceptor impurity $\varepsilon_m>\varepsilon_0$ and $t_m>0$. By solving the eigenvalues $E_i$ of $H_{0m}$, the density of states reads

$$D(E) = \sum_i \delta(E-E_i) = \lim_{\eta\to 0}\sum_i \frac{\eta}{\pi[(E-E_i)^2+\eta^2]}.$$ (21)

For atom number $N$=1000, strain $s$=0 and $\varepsilon_m=t_m$=3.0 eV, $D(E)$ is plotted in the upper panel of Fig. 3(b), indicating an impurity level above VB top at $E$=0.95 eV. For $\varepsilon_m=t_m$=2~4 eV, the energy of impurity level varies from -1.04 to -0.82 eV (the lower panel of Fig. 3(b)), which is above VB top and increases with $\varepsilon_m (t_m)$.

## VI. Photocurrent by ω+2ω dichromatic laser pulse

In semiconductor, photocurrent can be controlled through phase coherent control using one-photon excitation at frequency 2ω and two-photon excitation at frequency ω [30-35]. This technique has been applied on semiconducting molecular wires [36-38]. Here, the photocurrent generated by the interaction between LACC and ω+2ω dichromatic laser pulse was investigated. In the following simulation, the electric field of laser pulse reads

$$\mathbf{E} = \begin{cases} \mathbf{E}_0 \exp(-9(t-t_0)^2/t_0^2)[\sin(\omega t)+\frac{1}{2}\sin(2\omega t+\varphi)] & 0 \leq t < t_0 \\ \mathbf{E}_0 [\sin(\omega t)+\frac{1}{2}\sin(2\omega t+\varphi)] & t_0 \leq t \leq T-t_0 \;, \quad (22)\\ \mathbf{E}_0 \exp(-9(t-t_0+T)^2/t_0^2)[\sin(\omega t)+\frac{1}{2}\sin(2\omega t+\varphi)] & T-t_0 < t \leq T \end{cases}$$

corresponding to an additional laser pulse at frequency 2ω accompany with the pulse at frequency ω. The intensity of 2ω pulse is 1/4 times of ω pulse. The phase difference $\varphi$ controls the coherence of the two pulses. Gaussian profile, for which $t_0$=300 fs and $T$=1000 fs, was used for the starting and finishing stages of laser pulse. The electric field $\mathbf{E}$ is parallel to the direction of LACC, with a maximum intensity of $10^9$ W/cm$^2$ for the ω pulse (corresponding to E$_0$=8.68×10$^7$ V/m). As an example, the plot of $\mathbf{E}$~$t$ for $\hbar\omega$=0.2 eV and $\varphi$=0, π/2 and π is shown in Fig. 4(a). The corresponding vector potential reads $\mathbf{A} = -\int_0^t \mathbf{E} dt$. For a given laser wave packet with a vector potential $\mathbf{A}=\mathbf{A}(t)$, by the interaction described by Eq. (6) the Schrödinger equation for the evolution of every electronic states was simulated by the Runge-Kutta 4$_{th}$ order method. The photocurrent through the LACC was calculated by

$$\begin{aligned} \mathbf{I} &= -\frac{4e}{m}\sum_j <\Psi_j | \mathbf{p} | \Psi_j> \\ &= -\frac{4e}{m}\sum_{j,l,m} \mathbf{p}_{lm} <\Psi_j | c_l^+ c_m | \Psi_j> \end{aligned} \quad , \quad (23)$$

where the factor 4 comes from two spin degrees of freedom and two degenerate $p_y$, $p_z$ bands.

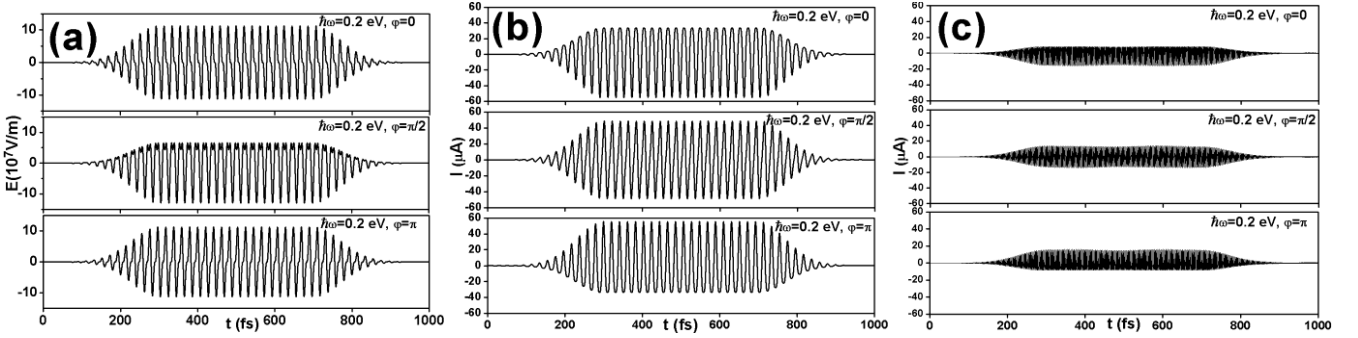

Fig. 4 (a) The plot of **E**~$t$ by Eq. (22) for $E_0=8.68\times10^7$ V/m and $\hbar\omega=0.2$ eV. The upper, middle and lower panels are for $\varphi=0$, $\pi/2$ and $\pi$, respectively. (b) The photocurrent **I** along the LACC for $\hbar\omega=0.2$ eV and the phase difference $\varphi=0$, $\pi/2$ and $\pi$, shown in the upper, middle and lower panels, respectively. (c) The photocurrent **I** along the LACC for $\hbar\omega=2.0$ eV and the phase difference $\varphi=0$, $\pi/2$ and $\pi$, shown in the upper, middle and lower panels, respectively.

For $\hbar\omega\ll E_g$, the photon energy is far off resonance and the electronic ground and excited states are nonadiabatically coupled by the dynamic Stark effect [36]. Simulations were performed for $\hbar\omega=0.1\sim0.5$ eV and $\varphi=0$, $\pi/2$ and $\pi$. For arbitrary laser frequency $\omega$, photocurrent **I** conducts along the LACC and synchronously oscillates with the laser field. With the beginning or end of the laser pulse, the photocurrent is increased or decreased. As an example, for $\hbar\omega=0.2$ eV the photocurrent generated in the case of $\varphi=0$, $\pi/2$ and $\pi$ is shown in the upper, middle and lower panels of Fig. 4(b), respectively. For the phase difference $\varphi=0$ or $\pi$, the oscillation of photocurrent **I** is obviously asymmetric along the two opposite directions of LACC, while for $\varphi=\pi/2$ the oscillation of photocurrent **I** is symmetric. For $\hbar\omega$ gets close to $E_g$, the photon energy is near-resonance and the photocurrent is mainly generated by the coherent excitation. The upper, middle and lower panels of Fig. 4(c) present the photocurrent for $\hbar\omega=2.0$ eV and $\varphi=0$, $\pi/2$ and $\pi$, respectively. Like for $\hbar\omega\ll E_g$, for $\varphi=0$ or $\pi$ the oscillation of **I** is obviously asymmetric, while for $\varphi=\pi/2$ the oscillation is symmetric. In the case of near-resonance, the amplitude of **I** is smaller than that of far off resonance, because for small $\hbar\omega$ the period of the electrons pulled along the LACC by the electric field is larger and obtains more energy than for near-resonance. By the result, for $\varphi=0$ or $\pi$ the photocurrent in LACC can be applied as light-controlled signals.

## VII. Summary

In this work, the optical properties of one-dimensional LACC and its response to laser field were investigated. In the range of elastic deformation, a direct band gap of LACC varies from 1.58 to 3.86 eV. For arbitrary strain, a sharp peak in the absorption spectrum locates at transition from VB top to CB bottom. Then, LACCs were proposed to be a medium of continuously tunable laser. By mechanical pulling, the wavelength can be continuously tuned in a range of 785~321 nm. In the range of elastic deformation, the lifetime of CB bottom is about 2.3~1.9 ns, slightly shortening with increasing strain. Such a lifetime is long enough as a laser medium. Population inversion between VB top and CB bottom can be easily implemented. The electrons pumped to CB will quickly fall to CB bottom by nonradiative transitions in a time of several ps. Finally, by $\omega+2\omega$ dichromatic laser pulse photocurrents can be generated when $\hbar\omega$ is far off resonance or near-resonance. Such photocurrents in LACC can be applied as light-controlled signals.

**Acknowledgements**

This work was supported by the National Natural Science Foundation of China under Grant No. 11304239, and the Fundamental Research Funds for the Central Universities.